# Superposition of flux-qubit states and the law of angular momentum conservation.

*Alexey Nikulov,* Institute of Microelectronics Technology, Russian Academy of Sciences, 142432 Chernogolovka, Moscow District, Russia. E-mail: nikulov@iptm.ru

**It is shown that the assumptions on macroscopic quantum tunneling and on superposition of two macroscopically distinct quantum states of superconducting loop, considered as flux qubit by many authors, contradict to the fundamental law of angular momentum conservation and the universally recognized quantum formalism.**

A superconducting loop interrupted by one or three Josephson junctions is considered by many authors [1-22] as flux qubit. As the authors [1] write, such superconducting qubits are not only of considerable fundamental interest but also might ultimately form the primitive building blocks of quantum computers. The modern technology is able to make the superconducting circuits which would be used as a register of quantum computer [23] if such generic quantum properties as superposition of states and entanglement, which are more commonly associated with atoms, could be observed on macroscopic level. Qubit (quantum bit), main element of quantum computer, is a two-state quantum system which can be in superposition of states. The authors of papers [1-22] are fully confident that the superconducting loop, which they consider as flux qubit, can be in superposition of two macroscopically distinct quantum states. I should show that this confidence is based on a delusion. The interpretations [24,25] of numerous experimental results as experimental evidence of superposition of macroscopically distinct quantum states (SMDQS) [26,27] and macroscopic quantum tunneling (MQT) [28-30] are in an irreconcilable contradiction with the law of angular momentum conservation and do not agree with the universally recognized quantum formalism.

The flux qubit, considered in [1] and many other publications [2-22], is a loop interrupted by three Josephson junctions in which a persistent current (called super-current in [1]) $I_q$ circulates clockwise or anti-clockwise, when magnetic flux $\Phi_e$ inside the loop is not divisible $\Phi_e \neq n\Phi_0$ by the flux quantum $\Phi_0 = \pi\hbar/e \approx 2.07 \times 10^{-15} \, T \, m^2$. SMDQS and MQT is assumed at $\Phi_e = (n + \frac{1}{2})\Phi_0$, when two permitted states with quantum numbers $n' = n$ and $n' = n + 1$ have the same minimum energy [1]. The persistent current $I_q$ in these states has the same value $|I_{q,n}| = |I_{q,n+1}| = I_{q,1/2}$ but opposite direction [1], for example clockwise $I_{q,n+1} = I_{q,1/2}$ in the state $n' = n+1$ and anti-clockwise $I_{q,n} = - I_{q,1/2}$ in the state $n' = n$. There is important to accentuate that magnetic moment $M_m = I_qS$ and angular momentum of Cooper pairs $M_p = (2m_e/e)M_M$ have different directions in the two states $n' = n$ and $n' = n + 1$. At the values $I_{q,1/2} \approx 5 \, 10^{-7} \, A$ and loop area $S \approx 10^{-12} \, m^2$ of a typical flux qubits [15] their values equal approximately $|M_{m,n}| = |M_{m,n+1}| = I_{q,1/2}S \approx 0.5 \, 10^5 \, \mu_B$ and $|M_{p,n}| = |M_{p,n+1}| = (2m_e/e)I_{q,1/2}S \approx 0.5 \, 10^5 \, \hbar$. Where $\mu_B$ is the Bohr magneton and $\hbar$ is the reduced Planck constant. The $M_p$ value is macroscopic since angular momentum of all $N_s = Vn_s$ Copper pairs in the loop with a macroscopic volume $V$ have the same value determined of the quantum number $n'$:

$$\oint_l dl p = \hbar \oint_l dl \nabla\varphi = \hbar n' 2\pi \qquad (1)$$

where $p = \hbar\nabla\varphi = mv + qA$ is the momentum of Copper pair, describing with the gradient $\nabla$ of phase $\varphi$ of the wave function $\Psi_{Sc} = |\Psi_{Sc}|exp(i\varphi)$ in accordance with quantum formalism; $|\Psi_{Sc}|^2 = n_s$ is the density of the Copper pairs. The relation (1) results from the requirement that the complex wave function $\Psi_{Sc} = |\Psi_{Sc}|exp(i\varphi)$ must be single-valued at any point of the superconductor loop. It is also the quantization $mvr = \hbar$ postulated by Niels Bohr for electron velocity $v$ on atomic orbit with a radius $r$ as far back as 1913. In the case of superconductor the momentum $p$ in (1) should be attribute to a single Cooper pair since the period of the oscillations $I_q(\Phi/\Phi_0)$ and all others corresponds to the charge $q = 2e$, $\Phi_0 = 2\pi\hbar/q = \pi\hbar/e$. But the quantum number $n'$ should be attribute to all $N_s = Vn_s$ Copper pairs since the angular momentum of a single pair can not change individually. Just therefore superconductivity is macroscopic quantum phenomena [31].

Thus, the authors [1-22] and many others are sure that superposition of states with different angular momentum is possible. Just the angular momentum but no its projection since the persistent current circulating in the plane loop can induce $M_m = SI_q$ and $M_p = (2m_e/e)M_M$ only in one z-direction, perpendicular to the loop plane, Fig.1. Therefore the magnetic moment and angular momentum should be considered as one-dimensional in this case. The loop, in contrast, for example, to atom or electron, is no central-symmetrical system. Therefore the quantum formalism of angular momentum and spin states of atom, electron and other elementary particle can not be applied to the loop. The application by many authors [1,4,5,17,22,24] of the spin- 1/2 formalism to the flux qubit misleads. The superposition of the z-projection eigenstates

$$|\Psi> = \alpha|\uparrow> + \beta|\downarrow> \qquad (2)$$

with $\alpha = 1/\sqrt{2}$ and $\beta = \pm1/\sqrt{2}$ presupposes the eigenstate of the x-projection

$$|\Psi> = |\uparrow> \text{ or } |\Psi> = |\downarrow> \qquad (3)$$

Therefore the formalism of superposition of angular momentum and spin states of central-symmetrical system does not contradict to the law of angular momentum conservation. Moreover it is based on this law [32].

In contrast to this the interpretations [24,25] of the experimental results [26-30] as an evidence of

SMDQS and MQT are in an irreconcilable contradiction with the law of angular momentum conservation. There is important to remind that quantum superposition and quantum tunneling do not require any external influence or any cause. The experimental "evidence" of MQT [29,30] is based on the temperature independence of the escape probability observed below a quantum-classical crossover temperature $T_{MQT} \approx 0.29$ K [29] or $0.1$ K [30]. The authors [29,30] and others assume that the potential barrier between two macroscopically distinct quantum states is overcome at $T > T_{MQT}$ because of thermal activation, i.e. because of chaotic interaction with environment, and at $T < T_{MQT}$ without any interaction. Thus, the authors of [29,30] and many other publication assume that the angular momentum of the quantum system can change on the macroscopic value $M_{p,n} - M_{p,n+1} \approx 10^5 \hbar$ without any interaction with other systems in defiance of the conservation law. According to the universally recognized quantum formalism it is impossible even for microscopic value $1/2 \hbar$.

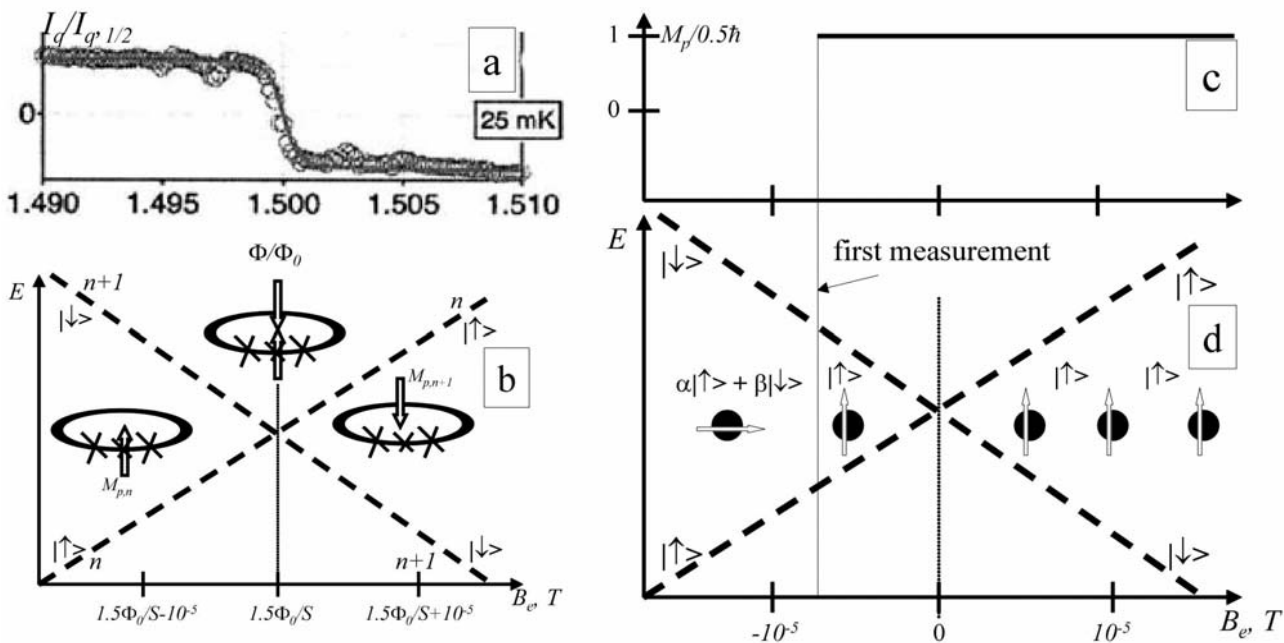

**Figure 1 | The angular momentum of the flux qubit consisting of a superconducting loop interrupted by three Josephson junctions and the spin – ½ of electron. a**, Normalized persistent current $I_q/I_{q,1/2}$ as a function of the external magnetic flux $\Phi/\Phi_0$ penetrating the flux qubit loop measured by the authors [34] at the temperature $T = 0.025$ K. Each open circle represents the results of many single-shot measurements taking an average. **b**, The experimental results collaborate that the quantum number $n'$ and the angular momentum $M_p$ of the flux qubit change with a variation of the externally applied magnetic field $B_e$ in a small interval from $B_e \approx 1.5\Phi_0/S - 10^{-5}$ T to $B_e \approx 1.5\Phi_0/S + 10^{-5}$ T. The probability of the $|\uparrow\rangle$ state with $n' = n = 1$ changes from $\approx 1$ to $\approx 0$ because of the increase of its energy $E$ and the energy decrease of the $|\downarrow\rangle$ state with $n' = n + 1 = 2$ in this $B_e$ interval. The average value of the persistent current $<I_q>$ and consequently of the angular momentum $<M_p> = (2m_e/e)<I_q>S$ measured at $B_e = 1.5\Phi_0/S$ equal zero since two states $|\uparrow\rangle$ and $|\downarrow\rangle$ have the same energy and consequently the same probability at $\Phi = (n+0.5)\Phi_0$. The assumption made by many authors that the change of the flux qubit angular momentum, $<M_p> \approx 0.5 \; 10^5 \; \hbar$ at $B_e \approx 1.5\Phi_0/S - 10^{-5}$ T, $<M_p> = 0$ at $B_e \approx 1.5\Phi_0/S$ T, $<M_p> \approx -0.5 \; 10^5 \; \hbar$ at $B_e \approx 1.5\Phi_0/S + 10^{-5}$ T, can occur because of macroscopic quantum tunneling and superposition of macroscopically distinct quantum states are in an irreconcilable contradiction with the law of angular momentum conservation. The measured $<M_p>$ value corresponds to no z-projection of the angular momentum but its full value since the persistent current induces the angular momentum in only direction perpendicular to the loop plane. **c**, According to the universally recognized quantum formalism the first measurement of the z-projection of a single electron in the x-projection eigenstate of spin-1/2, corresponding to the superposition $|\Psi\rangle = \alpha|\uparrow\rangle + \beta|\downarrow\rangle$ of the z-projection eigenstates ($\alpha = 1/\sqrt{2}$ and $\beta = \pm 1/\sqrt{2}$), should give result $M_p = 0.5\hbar$ or $M_p = -0.5\hbar$ with the same probability $|\alpha|^2 = 1/2 = |\beta|^2 = 1/2$. All posterior measurements of the z-projection should give the result $M_p = 0.5\hbar$ of the first measurement in conformity with the von Neumann's projection postulate. A different result observed without any interaction with environment and without measurement of an other spin projection could signify the violation of the law of angular momentum conservation. **d**, Because of the impossibility to imagine the superposition of states we force to use images of hidden variables. The arrow should not be interpreted as a real direction of electron spin which changes at the first measurement of the z-projection. The quantum formalism describes only phenomena, i.e. results of measurements, but no a real situation.

Measurements of the flux qubit persistent current [15,26,33,34] collaborate that the $M_m$ and $M_p$ direction changes with a small variation of the externally applied magnetic flux from $B_eS = \Phi_e = (n + ½ - \delta)\Phi_0$ to $\Phi_e = (n + ½ + \delta)\Phi_0$, corresponding to a field variation in the interval $2\delta B_e = 2\delta\Phi_0/S \approx 2 \cdot 10^{-5}$ T, Fig.1, at the typical values $\delta \approx 0.005$ and the loop area $S \approx 10^{-12}$ $m^2$ [15]. The SMDQS [26,27] and MQT [28-30] assumptions presuppose that this change can be without interaction with environment and even without any cause. But such causeless change is impossible even for a spin-1/2 projection. According to the habitual quantum formalism the wave function (2) describing the superposition of state should collapse to one of the eigenstates (3) at a measurement, for example, of the z-projection and all posterior measurements of the z-projection should give the same result, if interaction with environment is absent. The result measurements can not change irrespective of the $B_e$ variation in the narrow interval from $B_e \approx -10^{-5}$ T to $B_e \approx 10^{-5}$ T ($\mu_B B_e \approx \mu_B 10^{-5}$ T $\approx 0.9 \cdot 10^{-28}$ J $\approx k_B 10^{-5}$ K) or in a much wider interval, corresponding the energy $\mu_B B_e$ exceeding the temperature of measurement $\mu_B B_e > k_B T$ (the temperature of a typical flux qubit experiment $T = 0.01 – 0.03$ K [1]). Such change, for example from $|\uparrow>$ to $|\downarrow>$ would contradict to the law of the angular momentum conservation if it can occur without an interaction with environment. Therefore the universally recognized quantum formalism forbids a causeless change of the conserved quantities as energy, momentum, angular momentum and others at the consecutive measurements of the same parameter. The collapse of the wave function at measurement postulated in 1932 by von Neumann (von Neumann's projection postulate) [35] secures against a contradiction with the conservation laws at any observation.

The superposition of states is inconceivable without the von Neumann postulate since we can not see anything in two places at the same time or obtain contrary results, for example $|\uparrow>$ and $|\downarrow>$, at a single-shot measurement. The measurement, according to the quantum formalism, not only annihilates but also re-creates the superposition. For example, the x-projection measurement annihilates the superposition of the x-projection eigenstates and re-creates the superposition (2) of the z-projection eigenstates. The description of this superposition re-creation with help of a rotation operator [32]

$$U_y(\theta) = cos(\theta/2)+i\sigma_y sin(\theta/2) \quad (4)$$

(with $\theta = \pi/2$ or $\theta = -\pi/2$, $\sigma_y$ is a Pauli operator) accentuates once again that the spin-1/2 formalism can be applied only to a central-symmetrical system.

The two famous paradoxes proposed in 1935 year [36,37] have accentuated the paradoxical nature of the quantum phenomena described by superposition of state. The first paradox proposed by Einstein, Podolsky and Rosen [36] has revealed that this principle of quantum description contradicts to the local realism. In the version of the EPR paradox, proposed by David Bohm [38], spin-1/2 states of two particles $A$ and $B$ may be entangled

$$|\Psi> = \alpha|\uparrow>_A(r_A)|\downarrow>_B(r_B) + \beta|\downarrow>_A(r_A)|\uparrow>_B(r_B) \quad (5)$$

because of the law of angular momentum conservation. According to the von Neumann's projection postulate [35] a measurement of any spin projection of the particle $A$ located in a point $r_A$ should instantly change the spin state not only of this particle but also of the particle $B$

$$|\Psi> = |\uparrow>_A(r_A)|\downarrow>_B(r_B) \text{ or } |\Psi> = |\downarrow>_A(r_A)|\uparrow>_B(r_B) \quad (6)$$

located in a point $r_B$ irrespective of a distance $|r_A - r_B|$ between particles [25]. We should conclude that the quantum principle of state superposition presupposes the impossibility of a real existence of measured quantity (hidden variable) before the measurement or a possibility of non-local interaction, i.e. the "spooky action at a distance" which Einstein refuted. Thus, the fundamental difference of quantum bit from classical beat should be connected with violation of the local realism [39]. The numerous assumptions about the flux qubit should presuppose also violation of macroscopic realism. As it is written in [1,14] Anthony Leggett was first who proposed in the 1980s that a superconducting loop containing a Josephson tunnel junction could exist in a superposition of two macroscopically distinct quantum states. The title and essence of the paper [40] by A. J. Leggett and A. Garg emphasize that this assumption contradicts to macroscopic realism. This contradiction both with macroscopic realism and with noninvasive measurability at the macroscopic level [40] is beyond any doubt. But there is well-founded doubt [41] that the Leggett-Garg inequality [40] can reveal a contradiction of experimental results with macroscopic realism. In spite of similar mathematical forms of the Leggett-Garg [40] and Bell's [42,43] inequalities, they have an entirely different physical significance [41]. The famous no-go (no-hidden- variables) theorem by John Bell is based on a locality requirement [42], which is absent in [40]. The locality requirement is a decisive factor of the no-go theorem [42]. John Bell pointed out the hidden variable interpretation of David Bohm [44] as example of a non-local theory reproducing all observation prediction given by quantum formalism. He has shown [45] that von Neumann's no-go proof [35], which does not used the locality requirement, was based on an unreasonable assumption [46]. Bell has constructed [45] a hidden-variables model for a single spin – ½ that reproduces all predictions of measurement results given by the orthodox quantum theory using superposition of states [46]. This Bell's disproof of the von Neumann no-go theorem means that no experimental results obtained on a single two-state system can contradict to realism and give evidence of state superposition. Therefore it is strange that many authors [1-22] venture to interpret experimental results [6-10,15-21,27,28] obtained on superconducting loop and other systems with two states as an evidence of superposition of macroscopic states.

The second paradox, proposed in 1935 [37] is well known as "Schrodinger cat". It emphasizes that quantum tunneling and superposition express the absence of determinism in quantum description and it may be in quantum world. Since all cats, which we know, are macroscopic many authors [25,47-49] associate the "Schrodinger cat" paradox with the problem of macroscopic quantum phenomena. But anyone should easy understand that nothing in the Schrodinger paradox could depend on size of the cat. Moreover, anyone should easy understand that nothing, except tragedy situation, could change in this paradox at substitution of cat, small flask of hydrocyanic acid and hammer for a recorder which can record the discharge of Geiger counter tube. The Schrodinger cat paradox should associate rather with quantum measurement problem than with macroscopic quantum phenomena. We can say nothing about the initial cause of the measurement result which we have obtained. The $\Psi$-function of the entire system considered by Schrodinger [37]

$$|\Psi_{Sh.cat}> = \alpha|at_{yes}>|Ge_{yes}>|fl_{yes}>|cat_{dead}> +$$
$$+\beta|at_{no}>|Ge_{no}>|fl_{no}>|cat_{liv}> \quad (7)$$

describes the entangled states of atom $|at_{yes}>$, $|at_{no}>$, Geiger counter tube $|Ge_{yes}>$, $|Ge_{no}>$, small flask of hydrocyanic acid $|fl_{yes}>$, $|fl_{no}>$ and cat $|cat_{dead}>$, $|cat_{liv}>$. The items $|fl_{yes}>|cat_{liv}>$, $|Ge_{yes}>|fl_{no}>$, $|at_{yes}>|Ge_{no}>$ and so on are absent in (7) since the Schrodinger paradox presupposes the cause – effect connection between the states of the small flask and the cat, the Geiger counter tube and the flask, the atom and the Geiger counter tube. When we will open the steel chamber with the cat, the $\Psi$-function (7) describing superposition of states should collapse. If we see, for example, that the cat is dead

$$|\Psi_{Sh.cat}> = |at_{yes}>|Ge_{yes}>|fl_{yes}>|cat_{dead}> \quad (8)$$

we can conclude that the cat is dead since the hammer has shattered the flask with poison. The hammer has shattered the flask since the Geiger counter tube has discharged. The Geiger counter tube has discharged since the atom has decayed. Till now each result had the cause. But nobody can say why the atom has decayed. Just this absent of the cause is described by superposition of states (2,5,7) and quantum tunnelling used first by George Gamow in 1928 for description of the alpha decay of a nucleus. The quantum formalism assumes that phenomena, which we observe, can be causeless. But a causeless change of angular momentum or other violation of conservation laws are considered to be inadmissible.

The EPR and "Schrodinger cat" paradoxes are very important for the problem of quantum computer [39]. The idea of quantum computation is based on the very paradoxical principle introduced by Schrodinger [37,50] who was motivated [51] by the EPR paradox [36]. "Verschrankung" (in the German original [50]) or "entanglement" (in the English translation [37]) is called also "EPR correlation" [52]. Einstein, Podolsky and Rosen precluded a possibility of the EPR correlation [36] because of its contradiction with the principles of locality or realism. This contradiction means that in order to make a real quantum computer we should go out of the local reality. Because of this paradox the quantum computer is the problem not only technological and physical, but first of all philosophical. The "Schrodinger cat" can be in superposition of states (7) according to the quantum formalism. We may unite the EPR [36] and Schrodinger [37] paradoxes and show that states of two "Schrodinger cats" can be entangled. This means formally that the "Schrodinger cat" is quantum bit according to the quantum formalism. In contrast to this the superconducting loop is not flux qubit contrary to the confidence of many authors [1-22].

This false confidence resulted from a logic of universality. Most scientists were always fully confident that fundamental physical laws should be universally applied in all cases and on all level of sizes. This confidence may be based on the assumption that the physical laws describe an unique reality with a common entity. Quantum mechanics, in contrast to other physical theory, describes no unique reality but only phenomena. This unique feature of quantum mechanics has provoked the debates of many years about interpretation of quantum phenomena and quantum description, in particular between Einstein and Bohr. Albert Einstein emphasized that the quantum theory had relinquished precisely what has always been the goal of science: "*the complete description of any (individual) real situation (as it supposedly exists irrespective of any act of observation or substantiation)*" [53]. In contrast to this Niels Bohr argued that the very desire to seek such a complete description is misguided and naive. Quantum theory would provide predictions concerning the results of measurements, but, unlike all previous theories, it is incapable of providing a full account of "how nature did it." Bohr stated: "*There is no quantum world. There is only an abstract quantum physical description*", the citation from [54]. The experimental evidences [55-57] of violation of the Bell's inequalities compel us to agree with the Bohr's doubt about a real existence of quantum world. We can not be sure that quantum phenomena reveal elements of an objective reality. But if quantum mechanics describes only phenomena we can not be sure that the same *abstract quantum physical description* should be universally applied to all quantum phenomena, both atomic and macroscopic [31]. Therefore the logic of universality may mislead concerning description of quantum phenomena.

The authors of the papers [2,3,12,13,58,59] consider flux qubit and other superconductor structure as an artificial-atom. It is beyond any doubt that the persistent current is observed because of the Bohr's quantization just as stationary electron orbits in atom. But quantum phenomena observed in superconducting ring give evidence of fundamental difference between application of the Bohr's quantization in these two cases [60]. The observations [61,62] reveal that the persistent current has clockwise or anti-clockwise direction in contrast to electron velocity on atomic orbit. This violation of symmetry between opposite direction because of the Bohr's quantization [63] may be connected with the challenges to the law of momentum conservation [60,64] observed in some phenomena [62,65-67].

It is important to note that the wave function describing the real density of the Cooper pairs can not collapse [31] in contrast of the Shrodinger wave function interpreted as description of a probability. The real density $|\Psi_{Sc}|^2 = n_s$ can not collapse because of our look. This difference in essence [64] means that the wave function describing quantum phenomena in superconductor can not be used for description of state superposition. Therefore an new additional wave function, which can collapse, was fabricated [24] for the description of the flux qubit state superposition. Conjectural phenomena described by this wave function must violate the law of angular momentum conservation. Therefore a possibility to observe such phenomena is absolutely doubtful.